\documentclass[review,preprint,12pt]{elsarticle} 
\biboptions{sort&compress} 
\pdfoutput=1
\usepackage{amssymb}

\usepackage{graphicx}
\usepackage{epstopdf}
\newlength{\figurehalflength}
\newlength{\figurelength}
\usepackage{comment}
\usepackage{array}
\usepackage[below]{placeins}

\usepackage{amsmath}
\usepackage{url}
\usepackage{pifont}

\usepackage{color}
\usepackage{soul}

\usepackage{booktabs}

\usepackage{textcomp}

\journal{Cluster Computing}

\begin{document}
\DeclareGraphicsExtensions{.png,.jpg,.tif,.pdf}

\begin{frontmatter}

\title{Sustainability-Aware Cloud Computing Using Virtual Carbon Tax}

\author{Fereydoun Farrahi Moghaddam, Mohamed Cheriet}

\address{}

\begin{abstract}
In this paper, a solution for sustainable cloud system is proposed and then implemented on a real testbed. The solution composes of optimization of a profit model and introduction of virtual carbon tax to limit environmental footprint of the cloud. The proposed multi-criteria optimizer of the cloud system suggests new optimum CPU frequencies for CPU-cores when the local grid energy mix or the cloud workload changes. The cloud system is implemented on a blade system, and proper middlewares are developed to interact with the blades. The experimental results show that it is possible to significantly decrease the targeted environmental footprint of the system and keep it profitable.   
\end{abstract}


\end{frontmatter}

\section{Introduction}
\label{intro}

We live in the age of information while energy is still among the most, if not the most, important issues. Datacenters are where these two important phenomena (information and energy) interact with each other.\footnote{From the physics point of view, data processing is nothing more than converting energy from one form to another (electricity to heat).} The Information Technology (IT) sector is growing rapidly, and its hunger for energy is swiftly increasing  as well \cite{Lannoo2013}\cite{Frei2013}. Therefore, while we are waiting for a revolution in the clean, safe, cheap, accessible, and renewable energy, we need to smartly deal with current energy production technologies considering their huge spectrum of negative impacts. 

Virtualization and cloudification are new trends to bring flexibility and manageability to on-demand IT resources and services. Ultimately, virtualization and cloudification could mainly improve the energy efficiency of the system. However, a smart resource management in a cloudified system is required to achieve a higher level of performance in terms of energy efficiency and also other sustainability indicators \cite{lettieri2012expeditious}\cite{andrae2010life}\cite{finnveden2009recent}\cite{humbert2005impact}\cite{jolliet2003impact} such as profitability, environmental compatibility (GhG emission\footnote{GhG emissions include $CO_2$, $CH_4 $, $NO_x$, and $SO_2$.} reduction \cite{james1994environment}\cite{alexander2007environmental}\cite{Arushanyan2014}), and social responsibility. 

There are many research efforts done on different aspects of a sustainable cloud. However,  they often target one aspect of the sustainability and ignore the other important features of a sustainable cloud. 

This research presents a new design for a sustainable cloud that targets high profitability and low environmental impacts, simultaneously. In this research, we are considering a network of datacenters. This network of datacenters hosts a uniform cloud which itself hosts the VMs. This configuration makes the location of VMs a variable of the system. Also, we are using the Dynamic Voltage Frequency Scaling (DVFS) \cite{von2009power} technology in the blades that adds another variable to the equation of the system. Since the sustainability problem has many indicators, it is a multi-objective optimization problem \cite{azapagic1999life}. 

Due to the high number of parameters, variables, and objectives involved in our optimization problem, solving the multi-objective optimization problem as a whole can be very time-consuming. In \cite{kessaci2011pareto}, a similar optimization problem is solved with a heuristic approach; however, the approach can still be very slow for big networks.
In \cite{fereydoun2014}, we proposed a solution to convert the multi-objective problem of sustainability to a single objective problem of profitability. This is done by introduction of a {\em Virtual Carbon Tax} (VCT). In this research, VCT concept is extended and used for other sustainability indicators.  

In the following sections, first, we will discuss the problem statement of this research. Next, we will define the system and formalism of the multi-objective optimization problem of the system. Then, we will introduce a rapid solution for the optimization problem. Then, we will explain the steps have been taken to build the system in a real platform. Finally, we will present the results of the tests and conclusions.

\section{Problem Statement and Related Work}
\label{sec:literature}

The problem of cloud sustainability is the subject of many researches in recent years. However, it is very difficult to consider all the aspects of cloud sustainability in one work. This section highlights the strong and weak points of some of these studies, and next section proposes a design to overcome some of the issues of these researches. 

Various work have been done on energy awareness in distributed data centers. In \cite{Garg2011}, a two-level job scheduler for HPC workloads was proposed. In their research, they considered an average electricity mix and price for a set of geographically distributed data centers. In addition, they assumed that CPUs support dynamic voltage and frequency scaling (DVFS\footnote{DVFS is important because there is a nonlinear relation between CPU energy consumption and CPU-core frequency in contrast to linear relation between amount of executed instructions and CPU-core frequency. Therefore, a change in frequency have much higher impact on energy consumption than on the amount of executed instructions.}). Several greedy algorithms were included to reduce the carbon footprint and increase the profit while meeting the required QoS. They concluded that carbon footprint can be reduced with negligible fall in the profit.  Although the work is interesting, it suffers from various drawbacks. First of all, the electricity grid mix and price are approximated with their average values, while in reality, these values are highly variable and change even in an hourly scale. Moreover, in their DVFS-related optimization of the CPU-core frequency, they obtained a constant optimal frequency for each type of CPU-core that minimizes the energy consumption of each individual job. However, this unpenalized DVFS-based approach could result in low performance in terms of HPC requirements and QoS. 

Heuristic optimization has been also considered in many work for managing and scheduling jobs and applications. For example, \cite{Kessaci2011} used a multiobjective approach using a GA algorithm to schedule real HPC job traces on a distributed cloud. The solution was profit driven, and the cooling system was simply approximated using the Coefficient of Performance (COP) indicator. Similar to \cite{Garg2011}, the average values for electricity price and footprint (taken from EIA reports\footnote{\url{http://www.eia.doe.gov/cneaf/electricity/epm/table56a.html}}) were used. Also, the job deadlines were synthetically generated using the method proposed in \citep{Venugopal2008}. They compared their results with those of maximum resource utilization heuristic. The main drawback of the GA optimizers (and any other heuristic optimizer) is that they cannot consider the complex and dynamic configurations of free slots in their formalism. This condition highly simplifies the scheduling problem, and avoid maximum utilization of the resources.

The work \cite{zhang2010green} used DVFS to optimize the frequency of the running jobs to minimize the energy consumption of a system of heterogeneous cloud servers. However, they did not consider any other parameter which may play a role in the energy consumption, performance, and profit of the system. 

In another work, \cite{rizvandi2010linear}, it argued that a time slack will be produced after an optimum frequency is rounded up to the next possible CPU-core frequency (CPU-core frequencies are not continuous). Therefore, they breakdown the free slot in two pieces where the job will be run with two frequencies (maximum and minimum) instead of optimum frequency. However, they used a theorem\footnote{If $f_a$ and $f_b(>f_a)$ execute a task in $t_a$ and $t_b$, respectively. Then, $E(t_a)<E(t_b)$.} in their approach that is in inconsistency with the \cite{Garg2011} claim of having an optimum frequency for a given job with minimum energy consumption.

In \citep{Toporkov2011}, several free slot selection algorithms, such as algorithm based on Maximal job Price (AMP) and algorithm based on Local Price of slots (ALP), were considered in connection with a profit goal. However, the proposed scheduler, which runs on a synthesized simulator, does not consider the energy consumption or the carbon footprint associated with the operation of the network of datacenters.

In \cite{Qureshi2009}, a cost-aware request routing policy for Internet scale computing systems was introduced. The policy, which considers the variation of electricity price over time and location, preferentially maps the requests to those data centers that are cheaper. To model the cooling system, they simply used a constant PUE value. Also, the idle power consumption was assumed to be a percentage (65\%) of maximum power consumption. They showed that saving in electricity price is achievable if the electricity contracts are based on the actual power consumed and not the provisioned power ratings.

In summary, the problem of sustainable cloud is a complex problem with many parameters, variables, and outputs. In many studies, load balancing and DVFS are used as controlling tools for this problem. However, current solutions are lacking in addressing dynamic local energy mix and price \cite{Garg2011}\cite{Kessaci2011}\cite{zhang2010green}\cite{rizvandi2010linear}\citep{Toporkov2011}, total system energy efficiency \cite{Garg2011}\cite{rizvandi2010linear}\citep{Toporkov2011}, total system profit \cite{zhang2010green}\cite{Garg2011}\cite{rizvandi2010linear}, and the environmental impacts \citep{Toporkov2011}\cite{Qureshi2009}, or producing an efficient optimizer \cite{Kessaci2011}\cite{Garg2011}\cite{rizvandi2010linear}. In the following section, a cloud system is proposed to target profitability and environmental compatibility, simultaneously. The proposed system considers real-time local energy mix and price, and uses an efficient optimizer to minimize the environmental impacts of the system and to keep it profitable.

\section{Designing a Cloud System Optimized for Multi-Sustainability-Index}

\begin{figure}[!h]
  \centering
\includegraphics[width=8cm]{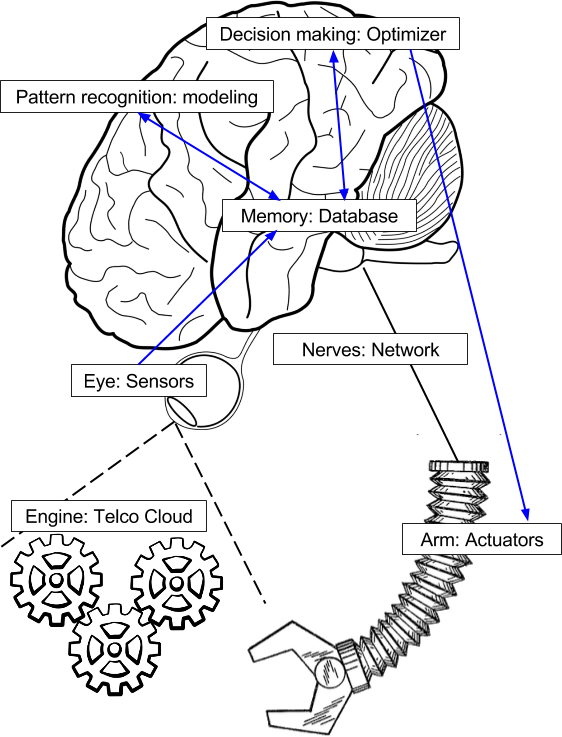}
\caption{A metaphoric view of the system.}
\label{fig:arch}
\end{figure}

Figure \ref{fig:arch} illustrates a metaphoric picture of the cloud system. This picture consists of five main parts: sensors, models, databases, optimizers, and actuators. The sensors are the eyes of the system. With them, the manager can collect the information form its own components and from its peripherals. These raw data is essential for the operation of the system. However, it is not enough for proper decision-making process. Therefore, several models are needed to calculate the necessary metrics such as environmental footprint and profit which are not directly measurable. The models act as learning entity of the system. For the purpose of accessibility and speed, these raw measures and calculated metrics are properly stored in databases. A database acts as the memory of the system. The optimizer uses this accessible information and calculates the best possible solution. The optimizer acts as the decision-support-system-unit of the system. Models, databases and optimizer represent the brain of the system. Finally, decisions need to be executed in the system by actuators that act as the arms of the system.

Having a successful sustainable cloud without any of these components seems impossible. In the following sections, the details of each of these components are described.

\subsection{Cloud Sysetm Sensors and Modeling} 

The key data requirements are values of blades-energy-consumption, blades-CPU-utilization, and blades-CPU-core-frequency. There is also a need for an energy model to calculate the energy consumption of the blades. This model is useful in the definition of the cost function of the optimization problem. The ``Building a Sustainable Cloud System'' section describes how raw data can be collected on a real system, and how energy consumption of blades can be modeled. 

Servers are not the only energy consumers of an ICT system. Cooling systems consume a significant amount of energy, which should be considered in any realistic model. A common method for considering the energy consumption of support in a data center, including the cooling system, is to use the Coefficient of Performance (COP) or Power Usage Effectiveness (PUE) factors:

\begin{equation}
\label{eq:etotal}
E_{\text{total}} = E_{\text{IT}} + E_{\text{support}}
\end{equation}
\begin{equation}
\label{eq:PUE}
\text{PUE} =  \frac{E_{\text{total}}}{E_{\text{IT}}}
\end{equation}


When the amount of consumed energy and the type of energy source is known, it is possible to calculate the real-time carbon footprint of the consumed energy by using the carbon emission factor of that particular source of energy. Other sustainability indicators can be calculated in the same way. 

In the real life, energy source of an electricity grid is a mixture of various energy source types, and this mixture changes over hours of a day. Therefore, a carbon emission factor for the grid is required to be calculated based on the participating energy sources and the amount of energy they are contributing in the grid at any time. Our model considers such a combination for the carbon footprint and other sustainability indicators of the grid.

Other metric that is required to be modeled is the profit of the cloud system. In this research, a flat rate has been considered for the cost of energy for the cloud that increases in the peak hours. In addition to the price of energy, there are already carbon taxes in place in some states and provinces that add an additional operational cost to the total cost of the services. There is also other related operational costs such as building rental, personnel, hardware/software investment, and network that need to be added to the total cost of the services. Furthermore, corporation tax needs to be considered in order to calculate the net, realistic profit of the blades.

\subsection{Cloud System Optimization}

There are three major elements in a sustainable cloud: profitability, low environmental impact, and its social aspects. Considering social aspects of a cloud system is out of scope of this research. In the following, the optimization problem of the cloud system is described.

\subsubsection{Multi-Objective Sustainability Optimization Problem}
\label{MOSOP}

Having the profit and sustainability indicator model of the system calculated, we can define the optimization problem of the system as follows:

\begin{equation}
\begin{array}{lll}
\min\quad && J = (-P(b,f),S(b,f)) \\
\textrm{subject to:} \\
b \in \mathbb{B}^m, && \mathbb{B} = {B_1, B_2, \cdots, B_n} \\
f \in \mathbb{F}^n, && \mathbb{F} = {F_1, F_2, \cdots, F_k} \\
cpu_i(b,f) \leq 1, && i=1, 2, \cdots, n \\
mem_i(b,f) \leq mem_{max}, && i = 1, 2, \cdots, n \\
net_i(b,f) \leq net_{max}, && i = 1, 2, \cdots, n  \\
\end{array}
\label{eq:12}
\end{equation}

Where, $P$ represent the profit of the system and $S$ represent the vector of other sustainability indicators of the system\footnote{In this research $S_1$ represent carbon footprint and $S_2$ represent the $SO_2$ footprint. Other sustainability indicators which appear in the $S$ vector are: $NO_x$ footprint, and fuel consumption indicator.} \cite{begic2007sustainability}: 
$$S(b,f) = \{S_1(b,f), S_2(b,f), \cdots , S_r(b,f)\}$$
$\mathbb{B}$ represents the set of available blades in the system in all datacenters. $\mathbb{F}$ represents the set of available frequencies for each blade. $cpu_i$, $mem_i$, and $net_i$ parameters ensure that the blades are not utilized more than their capacities. $n$, $m$, and $k$ represent the number of blades, the number of VMs, and the number of available CPU-core frequencies, respectively.

This problem is a multi-objective type of problem. Therefore, there are multiple acceptable solutions for this problem. Heuristic algorithms such as Genetic Algorithm (GA) can be used in order to produce the set of acceptable solutions of this problem (Pareto-front points in the criteria space). Especially, for network of datacenter type of problems, Multi-Level Grouping Genetic Algorithm (MLGGA) \cite{farrahi2014-mlgga} can be used which is proven to be efficient on this type of problems \cite{fereydoun2014}. The MLGGA algorithm is described in the following section.

However, regardless of efficiency of the MLGGA on this type of problems, heuristic algorithms are by nature slow algorithms, and they are especially much slower when it comes to multi-objective type of problems. For the manager of the system, it is very important to make quick decisions for the load balancing of the VMs and tuning the parameters of the system such as CPU-core frequencies of the blades. Therefore, in the following sections, a Virtual Carbon Tax is presented to convert the multi-objective problem to a single-objective problem.

\subsubsection{MLGGA Optimizer}

MLGGA algorithm is an extension of GGA algorithm \cite{falkenauer1992grouping}. The GGA algorithm is designed to help with bin packing problems in which normal GA algorithm is not efficient. In the GGA, the genetic representation is an array of groups (bins). The crossover and mutation in the GGA are designed to keep some group (bin) configurations from the parent genes. Although GGA works well on bin packing problems, it is not efficient on two level bin packing problems. A two level bin packing problem is a bin packing problem where the bins themselves are located in other bins. An example for bin packing problem is server consolidation for VMs in a datacenter. An example for two level bin packing problem is server consolidation for VMs in a network of datacenters. The MLGGA is designed to solve two or higher level bin packing problems. The genetic representation in MLGGA is an array of level-two or higher groups (bins). The crossover and mutation in the MLGGA are designed to keep some level-two or higher group (bin) configurations from the parent genes. 

\subsubsection{Virtual Carbon Tax Policy}
\label{VLT}

\begin{figure}[h]
  \centering
\includegraphics[width=9cm]{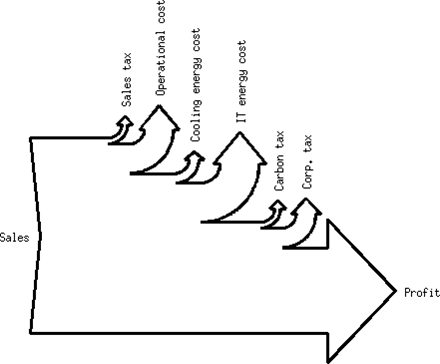}
\caption{Sankey diagram of profit of a cloud system.}
\label{fig:vct1}
\end{figure}

\begin{figure}[h]
  \centering
\includegraphics[width=14cm]{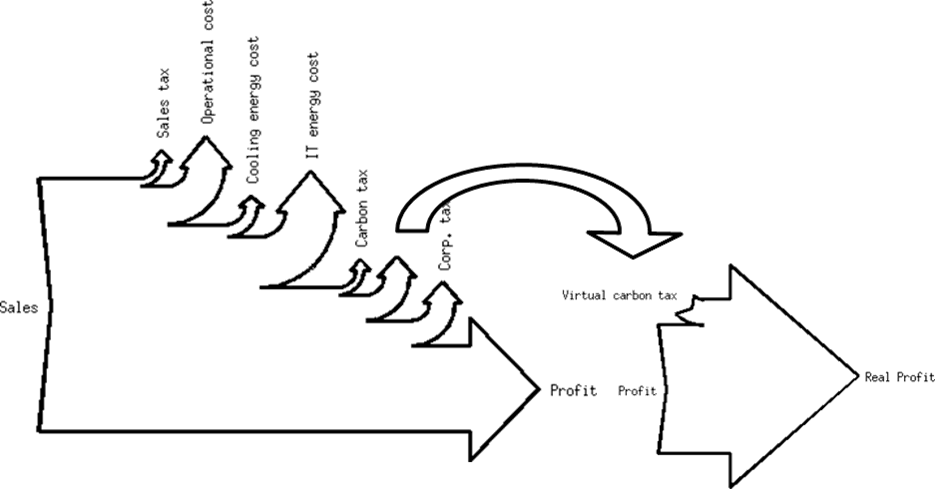}
\caption{Sankey diagram of profit of a cloud system under application of a virtual carbon tax.}
\label{fig:vct2}
\end{figure}

Virtual Carbon Tax (VCT) \cite{fereydoun2014} is a carbon tax that is virtually applied to the carbon footprint of the cloud system. This carbon tax does not exist in the real world. Therefore, the amount of the carbon tax is a part of the net profit of the cloud business as it is illustrated in the Figures \ref{fig:vct1} and \ref{fig:vct2}. Virtual carbon tax creates an illusion of a real tax for the optimizer. This illusion motivates the optimizer to reduce the carbon footprint of the cloud system to avoid a penalty. If the cost function of multi-objective optimization problem was described by $min(-profit(x),carbon(x))$, the new cost function is described as $min(-profit(x)+\sum_{i \in \mathbb{D}}{VCT_i * carbon_i(x)})$ which is a single objective function. Here, $x$ represents the variable vector of the system, and $\mathbb{D}$ represents the set of datacenters.


Similar to VCT, Virtual Sulfur Tax (VST) and other virtual taxes can be defined for other GHG footprint or other resource consumptions. In this paper, these virtual taxes are used to convert the multi-objective optimization problem to a single objective optimization problem. The single objective will be the virtual profit of the system. This will significantly speed up the process of optimization which is very necessary for this kind of applications. The new optimization problem after application of virtual taxes is as follows:

\begin{equation}
\begin{array}{lll}
\min \quad && J = (-\text{VP}(b,f)) \\
\textrm{subject to:} \\
b \in \mathbb{B}^m, && \mathbb{B} = {B_1, B_2, \cdots , B_n} \\
f \in \mathbb{F}^n, && \mathbb{F} = {F_1, F_2, \cdots , F_k} \\
cpu_i(b,f) \leq 1, && i=1, 2, \cdots , n \\
mem_i(b,f) \leq mem_{max}, && i = 1, 2, \cdots , n \\
net_i(b,f) \leq net_{max}, && i = 1, 2, \cdots , n  \\
\end{array}
\label{eq:13}
\end{equation}

Where, $\text{VP}$ represents the virtual profit of the system, which is described as follows:

\begin{equation}
\begin{array}{llll}
\text{VP}(b,f) & = & P(b,f) & \\
                 & - & \text{VCT} \times C(b,f) &\\
                 & - & \text{VST} \times S(v,f) &\\
                 & - & \text{VNT} \times N(v,f) &\\
                 & - & \text{VFT} \times I(v,f) &\\
                 & - & \text{VFT} \times O(v,f) &\\
                 & - & \text{VF}T \times B(v,f) &\\
\end{array}
\label{eq:13a}
\end{equation}

Where $\text{VCT}$, $\text{VST}$, $\text{VNT}$, $\text{VIT}$, $\text{VOT}$, and $\text{VBT}$ represent the virtual carbon, sulfur, nitrogen, iron, copper, and bauxite, respectively. $C(b,f)$, $S(b,f)$, $N(b,f)$, $I(b,f)$, $O(b,f)$, and $B(b,f)$ represent the carbon, sulfur, nitrogen, iron, copper, and bauxite indicators, respectively. $P(b,f)$ is defined as follows:

\begin{equation}
\begin{array}{llll}
\text{P}(b,f) & = & R(b,f) -(r_e * E(b,f) * PUE + O(b,f) + T(b,f))&\\
\end{array}
\label{eq:13a2}
\end{equation}

Where $R(b,f)$, $r_e$, $E(b,f)$, $O(b,f)$, $T(b,f)$ represents revenue of blade $b$ running at frequency of $f$, price of energy, blade energy consumption, operational costs (OPEX), and taxes, respectively. 

Note that the virtual taxes can be variable among datacenters. Therefore, the virtual profit equation can be rewritten as follow:

\begin{equation}
\begin{array}{lll}
\text{VP}(b,f) & = & P(b,f) \\
                 & - & \sum_{i\in\mathbb{D}}{\text{VCT}_i \times C_i(b,f)} \\
                 & - & \cdots \\
\end{array}
\label{eq:13b}
\end{equation}

Where $\mathbb{D}$ is the set of datacenters in the system. 

Now that we describe the multi-objective problem of the system and provide a methodology to convert it to a single objective (virtual profit), we can build the system and test the proposed methodologies on the real equipment.

\subsection{Other Components of the Cloud System}

With proper database tables and views definitions, it is possible to implement some of the models of the system inside the databases, and decrease the need for data processing outside the databases. This actions will significantly increase the speed of the manager.
\section{Building a Sustainable Cloud System}
\label{case}

\begin{figure*}[ht]
  \centering
\includegraphics[width=1.2\figurelength]{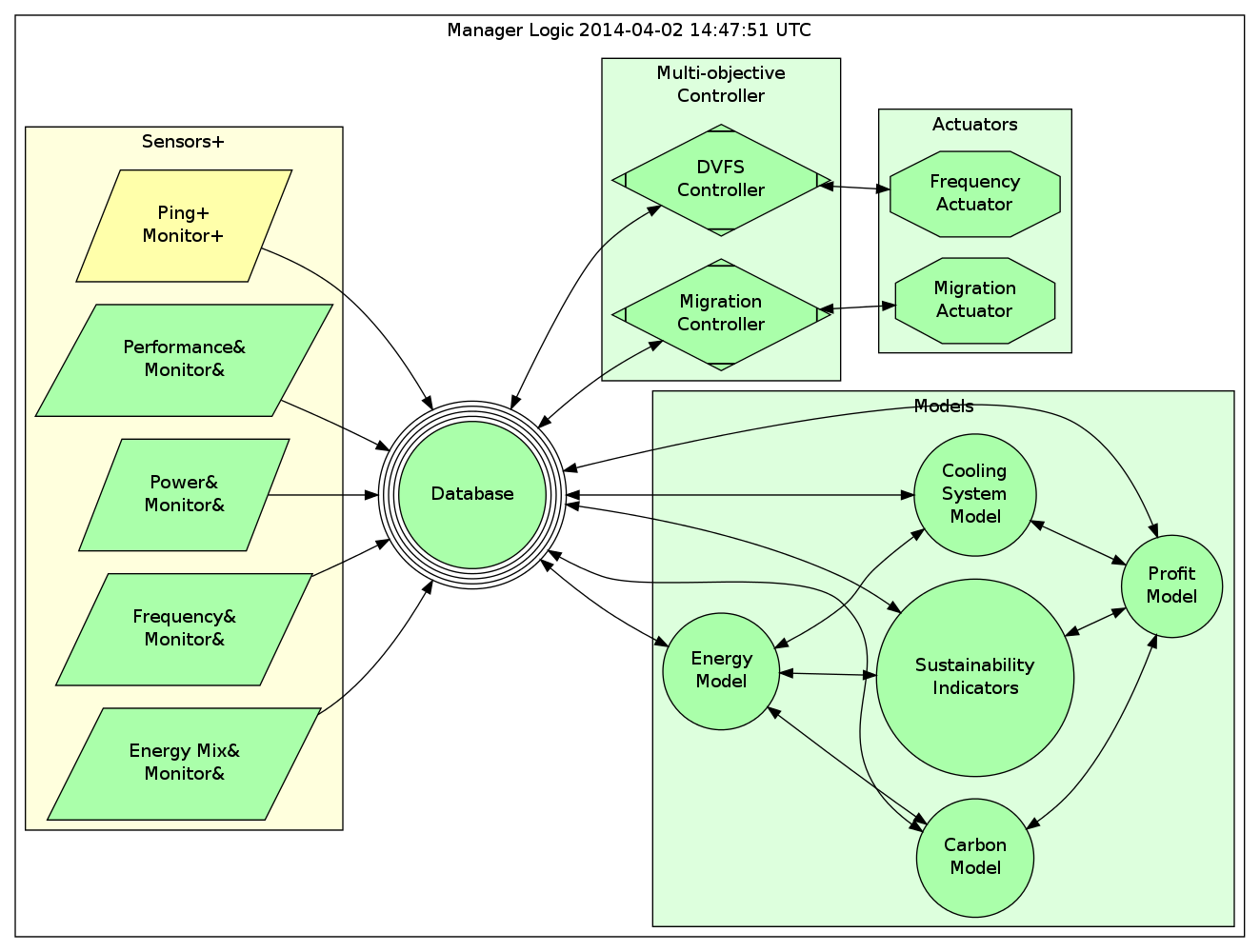}
\caption{System manager.}
\label{fig:manager}
\end{figure*}

Figure \ref{fig:manager} depicts a snapshot of the Graphical User Interface (GUI) developed for the cloud system manager in this research.

This figure describes the details of the components introduced in the system architecture with respect to the optimization problem defined in equation (\ref{eq:13}). As mentioned in the system architecture, one component of the system is the sensors. Sensors are responsible for collecting data from the system, and without these data, the manager will be blind. The manager uses these data for two reason. First, to be aware of the status of the system, and second, to evaluate the results of the modifications made to the system. For sensors, as well as other components of the system, service-oriented architecture is used. For example to collect the server utilization parameters, a RESTful web service is installed on the servers. This service records the data related to the server utilization of that particular server and stores them locally. Manager is then able to request for these data and also send commands to these RESTful services. Other sensor services which are implemented in the system are as follows: 1) power consumption sensors for IPMI supported blades, 2) sever CPU-core frequency monitor for DVFS supported blades, 3) region electricity mix data,\footnote{This is a service which scrape the data from Ontario hydro public website and provide it to the manager as a service.} and 4) ping server.

The manager uses the data provided by the sensor services and stores them in a central database\footnote{In this research, PostgreSQL database technology is used for storing the data.}. The collected data is provided in several database relational tables and views which will be used by the model component to calculate other metrics of the system such as energy consumption, environmental footprints, and profit.

\subsection{Real Testbed Setup}

In this implementation a Blade System (BS) is used to host the cloud. Every blade has an Intel CPU with six core, twelve threads (DVFS adjustable between 1.5Ghz and 2GHz). Our experiments show that the energy consumption of the blades are not achievable based on the typical cubic model used for the power consumption of a blade:

\begin{equation}
\label{eq:powerconsumption}
\begin{array}{lll}
&&P_{\text{Blade}_i} = \beta_i + \sum_{j\in\text{Cores}}{\alpha_i f_{\text{cpu}_j}^3}
\end{array}
\end{equation}

The power measurements from IPMI modules show that the power consumption of these blades does not fit on a cubic model. Therefore, for the purpose of this paper, a new model is manually fitted on the power consumption of blades based on the empirical data collected from the IPMI modules. The proposed equation (\ref{eq:ebspower}) estimates the power consumption of each blade with a small error margin.

\begin{equation}
\label{eq:ebspower}
\begin{array}{lll}
&P_{\text{Blade}_i} = &0.054 + 0.043 \big(\frac{u_{\text{Blade}_i}(t)}{u_{\text{Blade}_i}(t)+0.1}\big)+\\
&&0.012\big(\frac{f_{\text{Blade}_i}(t)-1.6}{0.4}\big)^{1.5} u_{\text{Blade}_i}(t)
\end{array}
\end{equation}

Where $P_{\text{Blade}_i}$, $u_{\text{Blade}_i}$, and $f_{\text{Blade}_i}$ represent the power consumption, cpu-utilization, and cpu-frequency of the $i^{\text{\tiny th}}$ blade (all the CPU-core frequencies are set to $f$), respectively. 

\begin{figure*}[ht]
  \centering
\includegraphics[width=1\figurelength]{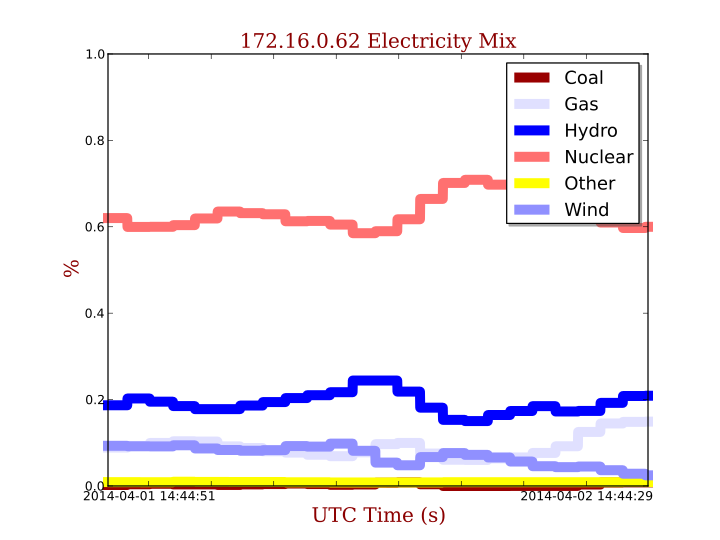}
\caption{Ontario energy mix.}
\label{fig:energymix}
\end{figure*}

Having the real-time electricity mix data (Figure \ref{fig:energymix}) from the electricity mix sensor and energy consumption model, we can calculate the sustainability indicators of the equipments and services in the use phase. Table \ref{tab:1} displays these indicators calculated for some energy sources \cite{begic2007sustainability}\cite{voss2001lca}\cite{afgan2002multi}\cite{gallego2010sustainability}\cite{afgan2000energy}\cite{evans2009assessment}.

\begin{table}  
\tiny
mean: here, top, bottom and on a separate page, respectively
\centering 
\begin{tabular}{l c c c c c c } 
\toprule 
& \multicolumn{3}{c}{Resource indicators} & \multicolumn{3}{c}{Environment indicators}  \\ 
\cmidrule(l){2-4} 
\cmidrule(l){5-7}
 & Iron & Copper & Bauxite & $CO_2$ indicator & $SO_2$ indicator & $NO_x$ indicator \\ 
Source of energy & (kg/GWh) &(kg/GWh) &(kg/GWh) & (g/kWh) & (g/kWh) & (g/kWh) \\ 
\cmidrule(r){1-1}
\cmidrule(l){2-7}
 Coal & 2310 & 2 & 20 & 815 & 13.8 & 0.8 \\
 Lignite & 2100 & 8 & 19 & 100 & 15 & 1 \\
 Gas  & 1207 & 3 & 28 & 400 & 4.5 & 0.4 \\
 Nuclear & 430 & 6 & 27 & 20 & 3 & 0.1 \\
 Solar & 6000 & 300 & 2350 & 200 & 17 & 0.45 \\
 Wind & 3700 & 50 & 32 & 30 & 4 & 0.12 \\
 Hydro & 2400 & 5 & 4 & 25 & 2.5 & 0.7\\

\bottomrule 
\vspace{0.05cm}
\end{tabular}
\caption{Sustainability indicators for sources of energy.} 
\label{tab:1} 
\end{table}

For the profit model, PUE is set to $1.2$, $R(b,f)$ is set to 10\textcent\  per core hour, $r_e$ is set to 8\textcent\  per kWh in normal hours and 16\textcent\  per kWh in peak hours, $O(b,f)$ is set to 2\textcent\  per core hour, and $T(b,f)$ is 10\% of the total profit. 

The result of calculated models will be stored in the databases in order to be available to the optimizer. As it was mentioned in the previous section, the goal of the optimization problem is to maximize the virtual profit of the system. The optimizer will use the information provided in the databases to create the optimization problem in each interval, and then will use a heuristic optimizer to find the best solutions of the system. Here, the variables of the optimization problem are the location of the VMs and frequencies of the CPU-cores of the blades. Regarding the small size of the experiment, in this research, the MLGGA \cite{farrahi2014-mlgga} is used in order to solve the optimization problem. For large size systems, greedy algorithms need to be used with the cost of having a lower accuracy of results in return of necessary speed \cite{fereydoun2014}.

Next step is to execute the result of the optimizer on the blades. Again, a series of services are installed on the blades regarding the migration of VMs and controlling the frequency of the CPU-cores. These services are the actuators of the system and without them the manager is not able to make any changes to the system. The migration actuator uses the KVM live migration capabilities in order to move the VMs from one blade to another blade. Our frequency actuator uses the ``cpufrequtils'' library to set the frequency of the CPU-cores on the blades.

In order to have a full picture of the system, the manager produces several reports, logs, and graphs which are accessible through the GUI that is specifically designed and implemented for this system.

\subsection{Experimental Results}

\begin{figure}[!h]
  \centering
\includegraphics[width=.8\figurelength]{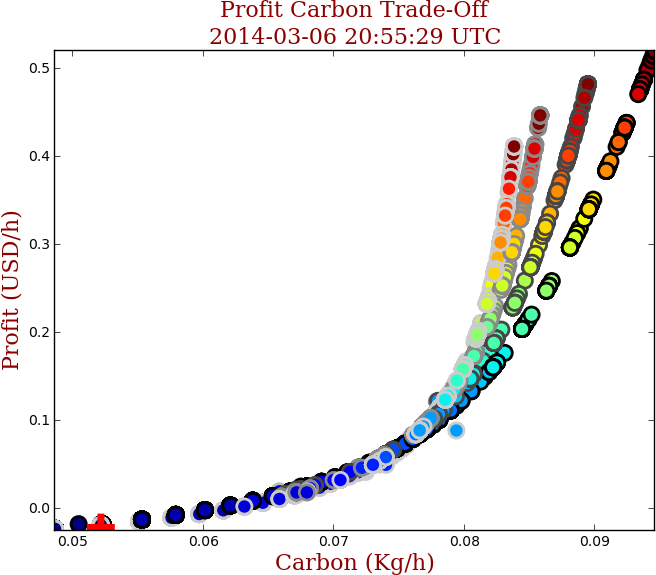}
\caption{Profit-carbon trade off.}
\label{fig:tradeoff1}
\end{figure}

In this research, a test system consisted of several physical servers were observed for its energy consumption, environmental footprints, and its profit. The system was tested under different loads, and the CPU-core frequencies of servers were set to arbitrary available frequencies. Figure \ref{fig:tradeoff1} shows the trade-off between profit and carbon in this system.

In this figure, different utilizations are colored from blue to red and different frequencies are depicted as points-borderline-grayness from light to dark. The current state of the system at the moment of production of the figure is depicted as a red plus sign. As it is shown, for similar utilization level and when the frequency decreases, the carbon reduction is faster than profit reduction. This characteristic is a good feature for this system. The optimizer can use this feature to reduce the carbon without impacting the profit as much. Worth noting that, all these points are the valid states of a blade and in a complete solution, any of these points could be part of the solution.

\begin{figure}[!h]
  \centering
\includegraphics[width=.8\figurelength]{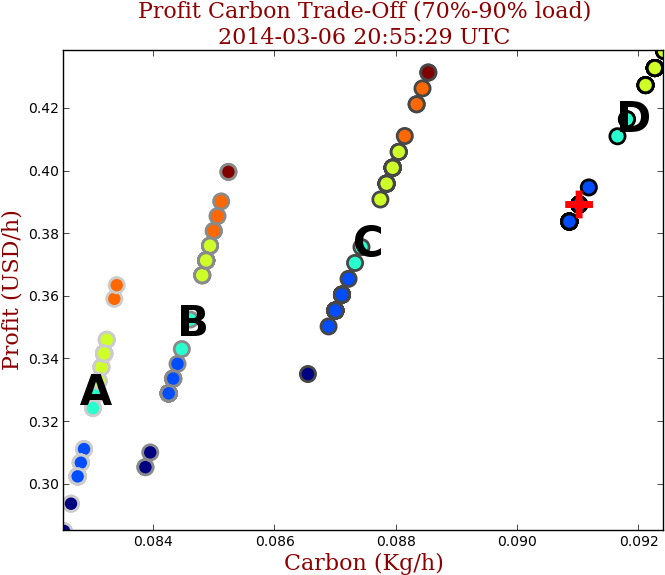}
\caption{Profit-carbon trade off (load: 70\%-90\%).}
\label{fig:tradeoff2}
\end{figure}

To have a better understanding of carbon-profit trade-off in the system, in Figure \ref{fig:tradeoff2}, only points near blade utilization = 70\%-90\% are shown and all other points are hidden. This graph is produced in the off-peak electricity price hours. There are four points that are marked with letters A, B, C, and D. These four points (described in Table \ref{tab:markdesc}) show the tradeoff between carbon and profit when utilization is 80\% for various frequencies.

\begin{table}
    \centering
\tiny
\begin{tabular}{|c|c|c|}
        \hline
        Marker & Utilization & Frequency \\ \hline
        A & 80\% & 1.6 GHz \\ 
        B & 80\% & 1.73 GHz \\ 
        C & 80\% & 1.86 GHz \\ 
        D & 80\% & 2.0 GHz \\
        \hline
    \end{tabular}
\caption{Markers description.} 
\label{tab:markdesc} 
\end{table}

\begin{figure}[!h]
  \centering
\includegraphics[width=.8\figurelength]{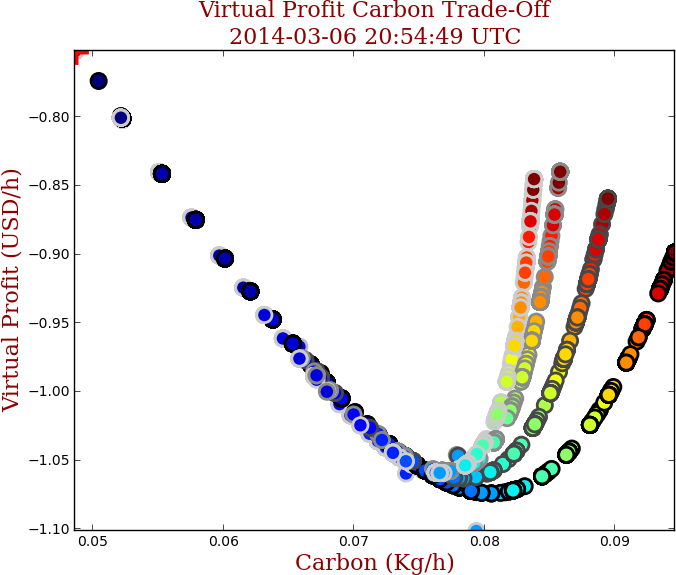}
\caption{Virtual profit-carbon trade off.}
\label{fig:tradeoff3}
\end{figure}

\begin{figure}[!h]
  \centering
\includegraphics[width=.8\figurelength]{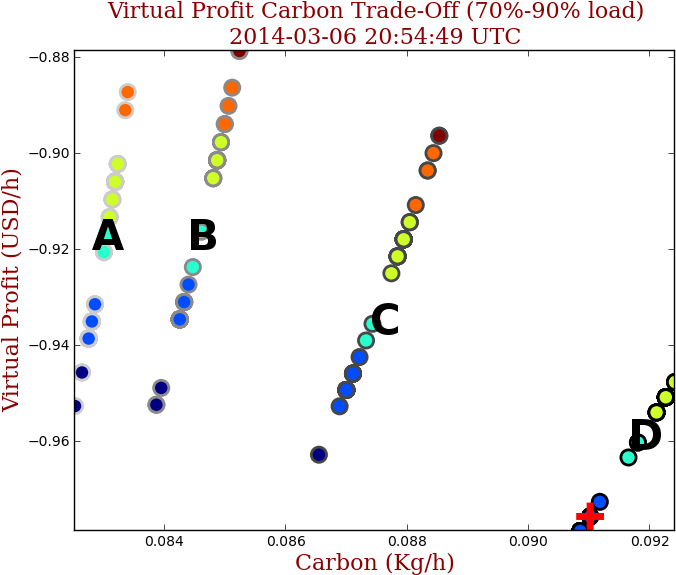}
\caption{Virtual profit-carbon trade off (load: 70\%-90\%).}
\label{fig:tradeoff4}
\end{figure}

As it is clear in the Figure \ref{fig:tradeoff2}, there is a clear tradeoff between profit and carbon of the system. Higher profit means higher footprint. To lower the footprint, the frequency of the CPU-cores is needed to be decreased which will lead to a decrease in profit. If the optimizer has only the profit as its single objective, the optimizer will choose the higher frequency, higher profit, and higher carbon. But this arrangement of the carbon and profit can be changed based on three parameters of the system. First, change in the energy price, i.e. peak hours. Second, change in energy mix of the grid (or switching to backup power generators which are usually with high environmental footprints). Third, application of the virtual taxes to artificially change the carbon-profit balance. As it is shown in the Figure \ref{fig:tradeoff3} and Figure \ref{fig:tradeoff4}, the optimizer need to lower the frequency of the CPU-cores in order to maximize the virtual profit which has a positive correlation with lower carbon footprint. It is worth noting that the virtual profit is only used for the optimizer process, and the real profit of the system is much higher than virtual profit of the system.

\begin{figure}[!h]
  \centering
\includegraphics[width=.8\figurelength]{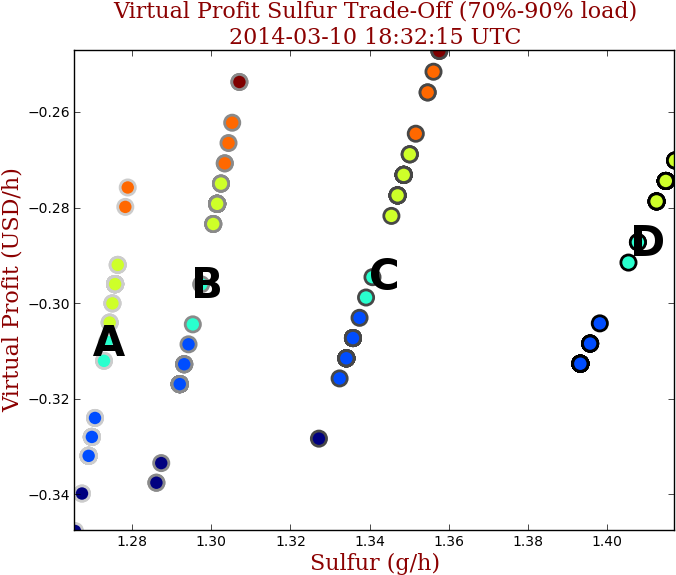}
\caption{Virtual profit-sulfur trade off (load: 70\%-90\%).}
\label{fig:tradeoff5}
\end{figure}

\begin{figure}[!h]
  \centering
\includegraphics[width=.8\figurelength]{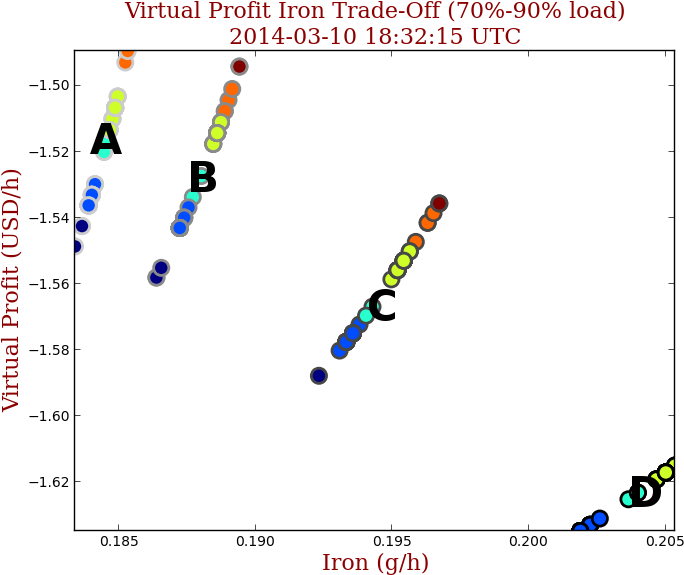}
\caption{Virtual profit-iron trade off (load: 70\%-90\%).}
\label{fig:tradeoff6}
\end{figure}

Two other sustainability indicators (sulfur and iron) and virtual profit are depicted in Figure \ref{fig:tradeoff5} and Figure \ref{fig:tradeoff6}, respectively. As shown, with a specific amount of virtual tax for sulfur and iron, the sulfur-virtual profit criteria space suggest higher frequencies while the iron-virtual profit suggest lower frequencies.
Each of Figures \ref{fig:tradeoff4}, \ref{fig:tradeoff5}, and \ref{fig:tradeoff6} represent the relation between virtual profit in one dimension with one of indicators of carbon, sulfur, or iron. However, the best frequency will be chosen by the optimizer from a hyperdimensional criteria space where the virtual profit is maximum.

Table \ref{tab:2} shows sample results of virtual profit optimization. The results are produced for Ontario energy mix at 2014-03-06 8pm and off-peak energy price. IP Multimedia Subsystem (IMS) modules were used to load the system to the desired level of utilization. For the revenue calculation, a simple model is used. A flat rate of 0.08 USD is used to calculate the revenue per GHz-hour utilization of the IMS modules. The optimization process chooses among points A, B, C, and D whichever virtual profit is higher. The amounts of virtual taxes were dynamically chosen by the manager to achieve the best results. Typical amount of virtual taxes are much higher than real taxes (e.g. a few dollar per kg carbon). The results show that large virtual taxes have a big impact on virtual profit (the value of virtual profits are negative\footnote{It means that if virtual taxes were real, the system was not profitable and therefore it was not sustainable.}) that leads to a significant reduction in negative sustainability indicators. However, as it is shown in the Table \ref{tab:2} (the real profit column), the system is still profitable. As it is marked in bold, the carbon and sulfur footprint and iron consumption are minimum when the respective virtual tax is applied. 

\begin{table*} 
\tiny
\centering 
\begin{tabular}{c c c c c c c c c} 
\toprule 
\multicolumn{3}{c}{Virtual tax applied} & \multicolumn{3}{c}{Sustainability indicators} & \multicolumn{2}{c}{Profit} & variable\\ 
\cmidrule(l){1-3} 
\cmidrule(l){4-6}
\cmidrule(l){7-8}
\cmidrule(l){9-9}
Carbon & Sulfur & Iron & Carbon & Sulfur & Iron & Real & Virtual &$f_{opt}$\\ 
(Boolean) & (Boolean) & (Boolean) & (g/kWh) & (g/kWh) & (g/kWh) & (USD/h) & (USD/h)& GHz\\ 
\midrule
 0 & 0 & 0 & 2.7                & {\bf 0.248}           & 0.168 & 0.417 & 0.417 & 2\\
 1 & 0 & 0  &  {\bf 2.025}   & 0.261          & 0.133 & 0.328 & -0.916 & 1.73\\
 0 & 1 & 0  & 2.7                & {\bf 0.248}           & 0.168 & 0.417 & -0.296 & 2\\
 0 & 0 & 1  & 2.138           & 0.33 & {\bf 0.126} & 0.286 & -1.518 & 1.6\\
 1 & 1 & 1    & 2.138           & 0.33 & {\bf 0.126} & 0.286 & -3.232 & 1.6\\

\bottomrule 
\vspace{0.05cm}
\end{tabular}
\caption{Experimental results.} 
\label{tab:2} 
\end{table*}
\section{Conclusion}
\label{sec:conclusion}

In this paper, a sustainable cloud system is introduced to maximize the profit and minimize the environmental footprint. First, the concept of such system has been presented as a metaphoric picture. Next, its associated multi-objective optimization problem has been defined in details, and the limitations in the path to find its possible solutions have been discussed. Then, a rapid mathematical solution has been proposed that can overcome the slow speed of the heuristic multi-objective optimizers. Finally, a complete prototype of the system has been implemented in a real world testbed for the proof of concept and to evaluate of the proposed methodologies.

The results of the implementation provide how the state of the system in the criteria space (profit-environmental footprint) changes based on the utilization and frequency chosen for the blades. The behavior of the states clearly shows the tradeoff between profit and environmental footprints. The newly proposed virtual environmental taxes, create a new criteria space between virtual profit and environmental footprint that can be completely different from the original criteria space. Using the virtual taxes, we were able to reduce the carbon footprint, other GHG emission indicators, and even other resource indicators, while keeping the system profitable.

\bibliographystyle{elsarticle-num}
\bibliography{bib}

\end{document}